\documentclass{JHEP3}

\usepackage{graphicx}
\usepackage{verbatim}
\usepackage{epsf}
\usepackage{latexsym}
\usepackage{amsmath,amsfonts,amssymb,amsthm}

\def\bseq{\begin{subequation}}  
\def\eseq{\end{subequation}}
\def\bsea{\begin{subeqnarray}}  
\def\esea{\end{subeqnarray}}

\newcommand{\bbox}{\lower.2ex\hbox{$\Box$}}

\newcommand{\beq}{\begin{equation}}
\newcommand{\eeq}{\end{equation}}
\newcommand{\bea}{\begin{eqnarray}}
\newcommand{\eea}{\end{eqnarray}}
\newcommand{\ena}{\end{eqnarray}}

\newcommand {\non}{\nonumber}

\renewcommand{\a}{\alpha}

\renewcommand{\l}{\lambda}

\newcommand{\Tr}{{\rm Tr}}

\newcommand{\be}{\begin{equation}}
\newcommand{\ee}{\end{equation}}

\title{\begin{center} 
$R$-symmetry and supersymmetry breaking\\
in $3D$ WZ models
\end{center}}
\author{Antonio Amariti$^{1,a}$ and Massimo Siani$^{1,b}$ 
\\ ~
\\
$^1$Dipartimento di Fisica, Universit\`a di Milano Bicocca\\
and \\
INFN, Sezione di Milano-Bicocca,\\ 
piazza della Scienza 3, I-20126 Milano, Italy\\
 \\
 ~~\\
  $^a$\email{antonio.amariti@mib.infn.it}\\
  $^b$\email{massimo.siani@mib.infn.it}
}
\abstract{
We analyze metastable supersymmetry breaking in $3D$ WZ
models. We study the regime of validity
of the perturbative computation 
for superpotentials with marginal and relevant couplings.
Lifetime of the metastable states in presence
of a triangular potential barrier is estimated.}

\begin{document}

\section{Introduction}

Three dimensional Chern-Simons (CS) supersymmetric gauge theories
recently raised a remarkable interest as candidates of field theory duals
in the $AdS_4/CFT_3$ correspondence. After the BLG
\cite{Bagger:2006sk}-\cite{Bagger:2007vi} and ABJM
\cite{Aharony:2008ug} models many CS gauge theories with
$\mathcal{N}=2$ supersymmetry have been conjectured to be dual to
$AdS_4$ backgrounds \cite{Martelli:2008si}-\cite{Aganagic:2009zk}.  In
this scenario the mechanisms of supersymmetry breaking have been still
rather unexplored.  In a recent paper \cite{Giveon:2009bv} the authors have
shown that a mechanism analog to the ISS takes place in three dimensional
massive SQCD with CS or YM gauge theories.  The low energy dynamics is
controlled by a Wess-Zumino (WZ) model. In four dimensions WZ models
have been useful laboratories for supersymmetry breaking
\cite{Ray:2006wk}- \cite{Intriligator:2008fe}, playing a crucial role
in the ISS mechanism \cite{Intriligator:2006dd}-\cite{Essig:2008kz}.

In this paper we analyze supersymmetry breaking in three dimensional
WZ models.  The WZ models studied in \cite{Giveon:2009bv} had relevant
couplings and the quantum corrections could be computed only after the
addiction of an explicit $R$-symmetry breaking deformation.  On the
contrary, a different solution to the problem of the computation of
quantum corrections in three-dimensional WZ models is given by
preserving an $SO(2)_R \simeq U(1)_R$ $R$-symmetry and by adding only
marginal deformations to the superpotential. The non supersymmetric
vacua turn out to be only metastable, since the marginal couplings
induce a runaway behavior in the scalar potential.  A property of
these models is that $R$-symmetry is spontaneously broken in the non
supersymmetric vacua.  As a general result it seems that in three
dimensions $R$-symmetry needs to be broken (explicitly or
spontaneously) for the validity of the perturbative expansion.

This paper is organized as follows. In section \ref{sezioneA} we review
the model of \cite{Giveon:2009bv} and the 
problems of the perturbative approach.  In section \ref{sezioneB} we
present a model with marginal couplings and long lifetime
metastable vacua.  
The general behavior of WZ models with marginal couplings
is studied in
Section \ref{sezioneC}. The regime of validity of the
perturbative approximation
in models with relevant coupling is 
discussed in section
\ref{sezioneD}. Finally we conclude in section \ref{sezioneE}.  
In appendix A we calculate the bounce action for a
three-dimensional triangle barrier potential. In appendix B we show a
formula to compute the one loop effective potential in every
dimension.

\section{Effective potential in $3D$ WZ models}
\label{sezioneA}

While a systematic study of supersymmetry breaking mechanisms in $3+1$
dimensions has been done, in $2+1$ dimension such an analysis
still lacks.  A recent step towards the comprehension of supersymmetry
breaking in $2+1$ dimensions has been done in \cite{Giveon:2009bv}. In this
section we briefly review their model and results.
\\
The theory is a WZ model with canonical K\"ahler potential
\begin{align}
  K = \Tr \left( M^\dag M + 
q_i^\dag q^i+ 
\tilde q_i^\dag \tilde q^i 
\right)
\end{align}
and superpotential
\be \label{Giveon:2009bvspot}
 W = h q M \tilde q +h \Tr\left(\frac{1}{2} \epsilon \mu M^2-\mu^2 M
 \right)
\ee
with an $U(N)\times U(N_F)$ global symmetry.   The
representations of the matrix valued chiral superfields $q$, $\tilde
q$ and $M$ are given in Table \ref{fieldsreps}.
\begin{table}
\begin{center}
\begin{tabular}{c|c|c|c}
&$q$&$\tilde q$ & $M$\\
\hline
$U(N)$&$N$&$\overline N$&$1$\\
\hline
$U(N_F)$&$\overline N_F$&$N_F$&$N_F^2$
\end{tabular}
\caption{Representation of the fields in
the model of \cite{Giveon:2009bv}}
\label{fieldsreps}
\end{center}
\end{table}
All the three dimensional couplings and fields in (\ref{Giveon:2009bvspot})
have mass dimension $1/2$, except $\epsilon$ which is adimensional. The 
model (\ref{Giveon:2009bvspot}) has supersymmetric vacua labeled
by $k=0,\dots,N$. At given $k$ the expectation values of the
chiral fields in the supersymmetric vacuum is
\be
M = 
\left(
\begin{array}{cc}
0&0\\
0&\frac{\mu}{\epsilon}\mathbf{1}_{N_F-k}
\end{array}
\right)
\quad
q \tilde q =
\left(
\begin{array}{cc}
\mu^2 \mathbf{1}_{k}&0\\
0&0
\end{array}
\right)
\ee
Moreover this model also has metastable vacua, 
in which the combination of the tree level and 
one loop scalar potential stabilizes the fields.
In the analysis of \cite{Giveon:2009bv} the authors studied 
the case of different values of $k$. Here we only
refer to the simplified case $k=N$. The vacuum is
\be\label{minimometa}
M = 
\left(
\begin{array}{cc}
0&0\\
0&X \mathbf{1}_{N_F-N}
\end{array}
\right)
\quad
q \tilde q =
\left(
\begin{array}{cc}
\mu^2 \mathbf{1}_{N}&0\\
0&0
\end{array}
\right)
\ee
where the field $X$ is a background field, 
stabilized, in this case,  by the one loop
effective potential. This potential is given by the Coleman-Weinberg
formula, that in three dimensions is \cite{Giveon:2009bv}
\be \label{CWa}
  V_{eff}^{(1)}=-\frac{1}{12 \pi^2} \text{STr}{|\mathcal{M}|^3} \equiv
  -\frac{1}{12 \pi^2} \Tr \left( {|\mathcal{M}_B|^3} -
  {|\mathcal{M}_F|^3} \right)
\ee
The cubic dependence on the bosonic and fermionic mass matrices ${\cal
  M}_B$ and ${\cal M}_F$ can be eliminated by expressing (\ref{CWa})
as
\be \label{CWa3}
V_{eff}^{(1)}=-\frac{1}{6\pi^2}\text{STr}\int_{0}^{\infty}
\frac{v^4}{v^2+\mathcal{M}^2}\text{d}v \ee
In appendix B we observe that (\ref{CWa3}) can be
generalized to every dimension.
\\
The superpotential that is necessary to calculate
the one loop corrections for the WZ model 
(\ref{Giveon:2009bvspot}) simplifies by expanding the fields 
around (\ref{minimometa}).
The fluctuations of the fields can be organized in
two sectors, respectively called 
$\phi_i$ and $\sigma_i$. The former represents the 
fluctuation necessary for the one loop corrections 
of the field $X$, while the latter parameterizes the
supersymmetric fields that do not contribute to 
the one loop effective potential.
We have 
\be \label{fluct}
q= \left(
\begin{array}{c}
\mu + \sigma_1\\
\phi_1
\end{array}
\right)
\quad
\tilde q^T= \left(
\begin{array}{c}
\mu + \sigma_2\\
\phi_2
\end{array}
\right)
\quad
M = 
\left(
\begin{array}{cc}
\sigma_3&\phi_3\\
\phi_4&X
\end{array}
\right)
\ee
The one loop CW is calculated by inserting (\ref{fluct}) in the
superpotential (\ref{Giveon:2009bvspot}). There are 
$N_F(N_F-N)$ copies of WZ models with superpotential
\be \label{eqn:Giveon:2009bv}
 W = \frac{1}{2} h \mu \epsilon X^2 - h \mu^2 X + h \mu \epsilon
 \phi_3 \phi_4 + h X \phi_1 \phi_2 + h \mu (\phi_1 \phi_3 + \phi_2
 \phi_4)
\ee
The tree level potential and the one loop corrections calculated from  
(\ref{eqn:Giveon:2009bv}) give raise to a non supersymmetric vacuum at
\footnote{For simplicity we consider the case of $k=1$.} 
\be \label{eqn:Xvev}
\phi_i=0 \ \ \ \   X \simeq \frac{\epsilon \mu}{b}
\ee
where $b=\frac{(3-2\sqrt 2)h}{4 \pi \mu}$ is a dimensionless
parameter. Thereafter we use a notation that 
makes clear the relevancy of the cubic three-dimensional couplings, by
rewriting (\ref{eqn:Giveon:2009bv}) as
\be
  W = \frac{1}{2} \epsilon \, m\, X^2 - f X + \epsilon \, m \, \phi_3
  \phi_4 + \frac{m^2}{f} X \phi_1 \phi_2 + m (\phi_1 \phi_3 + \phi_2
  \phi_4)
\ee
where we have defined
\be
  f \equiv h \mu^2 \qquad \qquad m \equiv h \mu
\ee
The new parameters of the theory,, $f$ and $m$,respectively have
mass dimension $3/2$ and $1$.
The $X$ field vacuum
expectation value is
proportional to $\epsilon f / (bm)$ and the expansion of the one-loop
potential near the origin is possible if the R-symmetry breaking
parameter $\epsilon$ satisfies
\be
  \epsilon \ll b
\ee
which expresses the condition $X\ll \mu$ of \cite{Giveon:2009bv}.

The perturbative expansion is valid if higher orders in the loop
expansion are negligible.  This last condition is satisfied when the
relevant coupling is small at the mass scale of the chiral fields
\be
h^2 \ll h X \iff \frac{m^4}{f^2} \ll \frac{m^2}{f} X
\ee 
This requirement imposes a lower bound on the $R$-breaking
parameter 
$\epsilon$
\be
  b \gg \epsilon \gg b \frac{m^3}{f^2}
\ee
and by using the definition of
$b=\frac{3-2\sqrt{2}}{4\pi}\frac{m^3}{f^2}$
\be
  \frac{3-2\sqrt{2}}{4\pi}\, \frac{m^3}{f^2} \gg \epsilon \gg
  \frac{3-2\sqrt{2}}{4\pi}\, \frac{m^6}{f^4}
\ee
The parameter $\epsilon$ cannot approach zero. In fact, in this case
the theory becomes strongly coupled and the effective potential cannot
be evaluated perturbatively. 

\section{Three dimensional WZ models with marginal couplings} 
\label{sezioneB}

Relevant couplings do not complete the renormalizable interactions
of a three dimensional superpotential.
In fact 
quartic marginal terms can be also added to a WZ model.
Here we study supersymmetry breaking in a renormalizable WZ
model with quartic marginal couplings and no trilinear interactions. 
We show that 
supersymmetry is broken at tree level and the perturbative
approximation is valid without any explicit $R$-symmetry breaking.
The three dimensional $\mathcal{N}=2$ superpotential is
\be\label{spotmarg1} W = -f X + h X^2 \phi_1^2 + \mu \phi_1 \phi_2 \ee
and the classical scalar potential is
\be \label{vtree} V_{\text{tree}}=|2 h X \phi_1^2-f|^2 +|2 h X^2
\phi_1 + \mu \phi_2|^2+|\mu \phi_1|^2 \ee
The chiral superfields have $R$-charges 
\be\label{Rcharges}
R(X)=2, \ \ \ \ R(\phi_1)=-1, \ \ \ \ R(\phi_2)=3
\ee
The $F$-terms of the fields $X$, $\phi_1$ and $\phi_2$ cannot be
solved simultaneously and supersymmetry is broken at tree level.  We study the
theory around the classical vacuum $\langle \phi_1 \rangle = \langle
\phi_2 \rangle =0$ and arbitrary $\langle X \rangle$. Stability
of supersymmetry breaking requires the computation of the one loop
effective potential for the $X$ field.  The squared masses of the
scalar components of the fields $\phi_1$ and $\phi_2$ read
\be \label{masses} m_{1,2}^2 = \mu^2 + 2 h \langle |X| \rangle
\left( h \langle|X| \rangle^3 +\eta f + \sigma \sqrt{f^2 + 2 \eta f
  h \langle |X| \rangle^3 + h^2\langle |X| \rangle^6 + \langle |X|
  \rangle^2 \mu^2} \right) \ee
where $\langle |X| \rangle$ is the vacuum expectation value of the 
field $X$ and $\eta$ and $\sigma$ are $\pm 1$.
These masses are positive for 
\be \label{bound}
\langle |X| \rangle < \frac{\mu^2}{4 f h}
\ee
In this regime the pseudomoduli space is tachyon
free and classically stable. Outside this region there 
is a runaway in the scalar potential.
The squared masses of the fermions in the superfields $\phi_1$
and $\phi_2$ are
\be \label{massesferm} m_{1,2}^2 = \mu^2 + 2 h \langle |X| \rangle
\left( h \langle|X| \rangle^3 + \sigma \sqrt{ h^2\langle |X|
  \rangle^6 + \langle |X| \rangle^2 \mu^2} \right) \ee
The two real combinations of the fermions of $X$ and 
$X^\dagger$ are the two goldstinos of the 
$\mathcal{N}=2 \rightarrow \mathcal{N}=0$ supersymmetry breaking.

The one loop effective potential is computed 
with the CW formula (\ref{CWa3}).
At small $\frac{X}{\sqrt{\mu}}$
the field $X$ has a negative 
squared mass
\be
m_{X=0}^2 \sim -\frac{f^2 h^2}{\mu}
\ee 
and the origin is unstable.

A (meta)stable vacuum is found if 
there is a minimum  
such that (\ref{bound}) is satisfied.
As long as the adimensional parameter $\frac{f^2}{\mu^3}$
is small  the effective
potential has a minimum at
\be \label{minimum}
\langle X \rangle \simeq \sqrt{\frac{\sqrt{2} \mu}{h}}
\ee
This imposes a bound on the coupling constant
\be\label{boundh}
h < \frac{\mu^3}{16 \sqrt{2}f^2}
\ee
since the scalar potential has to be tachyon free.

When (\ref{bound}) or (\ref{boundh}) are saturated
the classical
scalar potential (\ref{vtree}) has a runaway behavior.
Indeed, if we parameterize the fields by their
$R$-charges (\ref{Rcharges}), we have
%
%
\be  \label{parametrization}
X =\frac{f}{2 h \mu}e^{2\alpha} , \ \ \ \ \phi_1 =\sqrt{\mu}
e^{-\alpha} ,\ \ \ \ \ \phi_2 =-\frac{f^2}{2 h\sqrt{\mu^5}}e^{3\alpha}
\ee
and we get $F_X = F_{\phi_1} = 0$ and $F_{\phi_2} \rightarrow 0$ as $\alpha
\rightarrow \infty$.

\subsubsection*{Lifetime}
The decay rate of the non-supersymmetric state is proportional to the
semi-classical decay probability. This probability is proportional to
$e^{-S_B}$, where $S_B$ is the bounce action.  Here the lifetime of
the metastable vacuum is estimated from the bounce action of a
triangular potential barrier, since the two vacua are well separated
in field space and the maximum is approximately in the middle of them.

\begin{figure}
\begin{center}
\includegraphics[width=10cm]{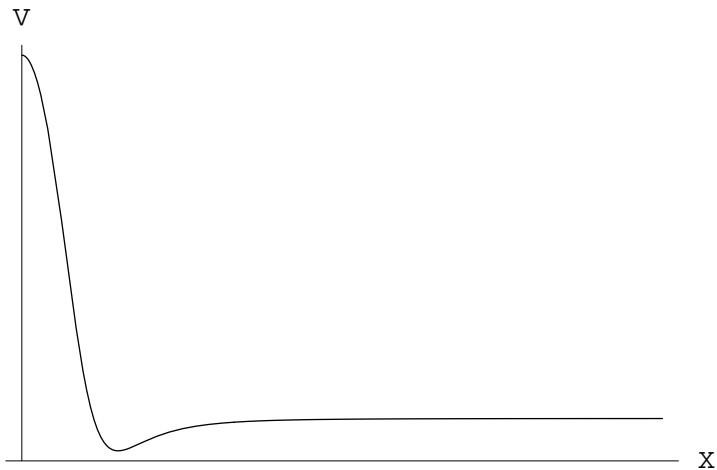}
\caption{One loop scalar potential for the one-loop
    validity region $X<\frac{\mu^2}{4 h f}$.  Over this value, we find
    a classical runaway.  At the origin the pseudomodulus has negative
    squared mass. The potential is plotted for $\mu=1$,
  $f=h=0.1$}\label{potenziale}
\end{center}
\end{figure}

The computation is similar to \cite{duncanjensen}, but in this case
we have to deal with a three dimensional theory. In the appendix
\ref{appendixB} we compute the bounce action for a triangular 
potential barrier in three dimensions.
We found that it is
\be
S_B \sim \sqrt{\frac{(\Delta \Phi)^6}{\Delta V}}
\ee
In our case we estimate the behavior of the potential barrier by
using the evolution of the scalar potential along the field $X$.  The
non supersymmetric minimum has been found in (\ref{minimum})
and the potential is $V_{min}=|f|^2$.
\\
The one loop scalar potential plotted in Figure \ref{potenziale}
is always increasing
between the metastable minimum and  
(\ref{bound}). When (\ref{bound})
is saturated there is a classical runaway direction, 
and the local maximum of the potential can be estimated to
be at $\langle X \rangle = \frac{\mu^2}{4 h f}$, where the  
potential is $V_{max}\sim 2|f|^2$.
After this maximum the potential starts to decrease and the field
$X$ acquires large values. There is not a local minimum, nevertheless
the lifetime of the non supersymmetric state can be estimated 
as in \cite{Shih:2007av,Franco:2006es}. Indeed, by using the parametrization
(\ref{parametrization}) of the fields along the runaway  
the scalar potential has the same value as $V_{min}$ for
$\langle X_{\text{R}} \rangle \sim \frac{\mu^2}{2 h f}$.

In the regime $\frac{f^2}{\mu^3} \ll h \ll 1$ 
the barrier is approximated 
to be triangular, and the gradient of the potential
is constant. The non supersymmetric state is near the origin of the 
moduli space, and the bounce action is
\be
S_B \sim \sqrt{\frac{\langle X_{\text{R}}\rangle^6}
{V_{\text{min}}}} \sim \sqrt{
\frac{1}{h}\left(\frac{\mu^3}{f^2}\right)^4}\gg
\left(\frac{\mu^3}{f^2}\right)^2
\gg 1
\ee

\section{The general case}\label{sezioneC}

In this section we consider the class of models with a single
pseudo-modulus $X$ which marginally couples to $n$ chiral
superfields $\phi_i$. As shown in \cite{Shih:2007av} for the class of
four-dimensional renormalizable and $R$-symmetric models, many general
features are worked out by $R$-symmetry
considerations. For the three-dimensional case, we find some
interesting properties concerning such models.  The perturbative
expansion is reliable under the weak condition that the coupling
constants are small numbers, i.e. one can use the one-loop
approximation and made higher loop corrections suppressed. The origin
of the moduli space is a local maximum of the one-loop potential, and the
pseudomodulus acquires a negative squared mass. Finally the scalar tree-level
potential exhibits runaway directions for every choice of the
couplings.

To deal with renormalizable three-dimensional WZ models, we consider
only canonical K\"ahler potential,
and superpotential of the type
\begin{align} \label{eqn:general}
  W = -f X + \frac12 ( M^{ij} + X^2\, N^{ij} ) \, \phi_i \, \phi_j
\end{align}
in which $R$-symmetry imposes the conditions
\be \label{eqn:NM}
  M^{ij} \neq 0 \Rightarrow R(\phi_i)+R(\phi_j) = 2 \qquad \qquad
  N^{ij} \neq 0 \Rightarrow R(\phi_i)+R(\phi_j) = -2
\ee

The conditions (\ref{eqn:NM}) could not be sufficient to uniquely fix
the $R$--charges. However in this case there exists a basis in
which there are both charges greater than two and charges lower than
two.

In a basis where the fields with the same $R$-charge are grouped
together, the $M$ matrix is written in the form
\be \label{eqn:M}
  M = \left(
  \begin{array}{cccccccc}
    & & & & & & & M_1 \\
    & & & & & & M_2 & \\
    & & & & . & & & \\
    & & & . & & & & \\
    & M_2^T & & & & & & \\
    M_1^T & & & & & & & \\
  \end{array}
  \right)
\ee
and similar for the $N$ matrix.
The scalar potential of this model can be written as
\begin{align}
  V_S = \left| -f + X N^{ij} \phi_i \phi_j \right|^2 + \left| M^{ij}
  \phi_j + X^2 N^{ij} \phi_j \right|^2
\end{align}
which we assume to have a one-dimensional space of extrema given by
\begin{align} \label{eqn:extrema}
  \phi_i = 0 \qquad \qquad X\, {\rm arbitrary} \qquad \qquad V_S =
  \left| f \right|^2
\end{align}
For general couplings, there can be other extrema and in particular
some lower local minima away from of the origin of $\phi$'s.
Furthermore for
some choices of the coupling constants at least some of
(\ref{eqn:extrema}) are saddle points. Here we work under the
hypothesis that this is not the case.

We show now that the effective potential always has a local maximum at
the origin of the pseudo-moduli space:
\begin{align}
  V_{eff}(X) = V_0 + m^2_X |X|^2 + {\cal O}(X^3)
\end{align}
and the $X$ field acquires a negative squared mass. We derive a
general formula for $m^2_X$ in the one-loop approximation by using
the Coleman-Weinberg formula
\bea
  V_{eff}^{(1)} &=& - \frac{1}{12\pi} {\rm S} \Tr |{\cal M}|^3 \non
  \\ &=& - \frac{1}{6\pi^2} \Tr \int_0^\Lambda \, dv \, v^4 \left(
  \frac{1}{v^2 + {\cal M}_B^2} - \frac{1}{v^2 + {\cal M}_F^2}
  \right) \label{eqn:CWc}
\eea
where ${\cal M}_B^2$ and ${\cal M}_F^2$ are, respectively, the squared
mass matrices of the bosonic and fermionic components of the
superfields of the theory
\begin{align}
\begin{split}
  {\cal M}_B^2 &= 
  (\hat M + X^2 \hat N)^2 - 2 f X \hat N \\
  {\cal M}_F^2 &= 
  (\hat M + X^2 \hat N)^2
\end{split}
\end{align}
where we have defined
\begin{align} \label{eqn:MNhat}
  \hat M \equiv \left(
  \begin{array}{cc}
    0 & M^\dag \\
    M & 0
  \end{array}
  \right) \qquad \qquad
  \hat N \equiv \left(
  \begin{array}{cc}
    0 & N^\dag \\
    N & 0
  \end{array}
  \right)
\end{align}
which take a form analogous to (\ref{eqn:M}).

Substituting the mass formulas into the Coleman-Weinberg potential
(\ref{eqn:CWc}) and expanding up to second order in the field $X$ we
find
\begin{align} \label{massimoculo}
  V_{eff}^{(1)} &= - \frac{1}{6\pi^2} \Tr \int_0^\Lambda \, dv \, v^4
  \, \frac{1}{v^2 + \hat M^2} \left( \frac{1}{1 + \frac{ X^2 \{ \hat
      M, \hat N \} - 2 f X \hat N }{v^2 + \hat M^2}} - \frac{1}{1 +
    \frac{ X^2 \{ \hat M, \hat N \} }{v^2 + \hat M^2} } \right) +
  \ldots \non \\ &= - \frac{2 f^2 X^2}{3\pi^2} \Tr \int_0^\Lambda \,
  dv \, v^4 \, \frac{1}{v^2 + \hat M^2} \, \frac{1}{v^2 + \hat M^2} \,
  \hat N \, \frac{1}{v^2 + \hat M^2} \, \hat N + {\cal O}(X^3) \non
  \\ &= -\frac{f^2 X^2}{2\pi^2} \Tr \int_0^\Lambda \, dv \, v^2 \,
  \frac{1}{v^2 + \hat M^2} \, \hat N \, \frac{1}{v^2 + \hat M^2} \,
  \hat N + {\cal O}(X^3)
\end{align}
where the last step follows after an integration by parts. From the
previous formula we note that the origin of moduli space is always a
local maximum, i.e. the pseudo-modulus always acquires a negative
squared mass at one-loop level. 
The vacuum cannot be at the origin, and we have to find it 
at $X\neq0$, where  $R$-symmetry is spontaneously
broken. 
The existence of this vacuum 
is not guaranteed by (\ref{massimoculo}), 
but it depends on
the couplings in (\ref{eqn:MNhat}) 

\vspace{1cm}

We show now that the models in (\ref{eqn:general}) have a runaway direction. In
four-dimensional theories if the $R$-charges of the superfields are
both greater than two and lower than two then the potential exhibits
runaway \cite{Carpenter:2008wi}. In three-dimensional renormalizable theories
(\ref{eqn:general}), conditions (\ref{eqn:NM}) state there are always
both superfields with $R$-charge greater than two and superfields with
$R$-charge lower than two. We parametrize the fields by their
$R$-charge $R(\phi_i) \equiv R_i$ as
\bea \label{parfields}
  &\phi_i = c_i \, e^{R_i\a} \non \\ &X = c_X \, e^{2\a}
\eea

The runaway behavior of the potential is analyzed by looking at the 
$R$-charges of $F$-terms. The $F$-terms with $R$ charges lower or equal 
to zero can be solved. All the non vanishing $F$ terms have charge greater 
than zero. They vanish only if $\a\rightarrow\infty$. 
By (\ref{eqn:NM}), the $F$-terms are
\be
\label{primo}
  F_X = -f + X N^{ij} \phi_i \phi_j = -f + N^{ij} c_X c_i c_j 
~~~~~~~~~~~~~~~~~~~~~~~~~
\ee
\be
\label{second}
~~~~~~~~~~~~~~  
F_{\phi_i}  = M^{ij} \phi_j + X^2 N^{ik} \phi_k = \left( \sum_j
  M^{ij} c_j + \sum_k N^{ik} c_X^2 c_k \right) \: e^{(2-R_i)\a}
\ee
The $F$-terms that are not zero in (\ref{second})
vanish when $\a \rightarrow \infty$, which implies
a runaway behavior 
for some of the fields in (\ref{parfields})

\section{Relevant couplings}\label{sezioneD}

In three dimensions there exist WZ models with relevant deformations
that can be perturbatively studied without the addition of explicit
$R$-symmetry breaking deformations.  Even if $R$-symmetry is not
explicitly broken in three dimensions, quantum
non supersymmetric vacua can appear out of the origin of the moduli
space. The vevs at which the vacua are found set the spontaneous
$R$-symmetry breaking scale which plays the same role as the
$\epsilon$ deformation in \cite{Giveon:2009bv}.
 
The models which exhibit explicit or spontaneous $R$-symmetry breaking 
can be perturbatively studied in three dimensions. The
former case has been analyzed in \cite{Giveon:2009bv}.  
We treat here the latter
case.  Assuming the $R$-symmetry breaking vacuum is near
the origin, we require that the pseudomodulus acquires a negative squared
mass at the origin of moduli space, i.e.there is a vacuum of the
quantum theory which spontaneously breaks the $R$-symmetry. 
This happens if not all the $R$-charges take the values $R=0$ and $R=2$.
This result was shown in four dimensions in \cite{Shih:2007av} and can
be analogously demonstrated in three dimensions.  We classify these
models in two subclasses.

In the first class we identify the models without runaway behavior,
i.e. all the charges are lower or equal to two.  There can be a regime
of couplings in which supersymmetry is broken in non $R$-symmetric
vacua. The vev of the field that breaks $R$-symmetry introduce a scale
which bounds the perturbative window for the relevant couplings.

In the second class, that we consider in the following, there are
runaway models. They have $R$-charges lower and higher than two.
Under the assumption of a hierarchy on the mass scales, we distinguish
two possibilities. Some of these models flow in the infrared to models
with only marginal couplings, that have been treated in sections
\ref{sezioneB} and \ref{sezioneC}.  The other possibility is that the
effective descriptions of these theories share marginal and relevant
terms in the superpotential.  In both cases a perturbative regime is
allowed.

\subsection{A model with relevant couplings}

Consider the superpotential 
\bea \label{spotrel}
W &=& \l X \phi_1 \phi_2 - f X +\mu \phi_1 \phi_3 + \mu \phi_2 \phi_4 
\nonumber \\
&+&m \phi_1 \phi_5 + m\phi_3 \phi_5
\eea
The first line corresponds to the ISS low energy superpotential,
while the second line of (\ref{spotrel})
asymmetrizes the behavior of $\phi_1$ and $\phi_2$.
If the mass term $m$ is higher than the other scales of the theory
($m^2 \gg \mu^2$ and $m^2 \gg f^{4/3}$) we can integrate out the 
second line of (\ref{spotrel}), and obtain
\be \label{spotmarg}
W = h X^2 \phi_2^2 - f X + \mu \phi_2 \phi_4 
\ee
where $h=\frac{\l}{4\mu}$ is a marginal coupling.
The perturbative analysis of this model is
now possible, with the only requirement that
$h \ll 1$.

The superpotential (\ref{spotmarg}) is identical to
(\ref{spotmarg1}). 
This example shows that 
in three dimensions models with cubic 
couplings in the UV can flow to theories with quartic
terms in the IR, which are perturbatively accessible.

\subsection{Models with relevant and marginal couplings}

A model with relevant couplings can  
flow to a model with both marginal and relevant couplings.
For example if we take the superpotential
\be
W = -f X+ \l_5 X \phi_1 \phi_5 - \frac{m}{2} \phi_5^2 
+\mu\phi_1\phi_2+\l X\phi_3^2+\mu\phi_3\phi_4 
\ee
and we study this model in the regime $m^2\gg\mu^2$, $m^2\gg f^{4/3}$,
we can integrate the field $\phi_5$ out. The effective theory
becomes
\be
W = -fX+h X^2\phi_1^2+\mu\phi_1\phi_2+\l X\phi_3^2+\mu\phi_3\phi_4 
\ee
where $h=\frac{\lambda_5^2}{m}$.
This model preserves $R$-symmetry and the charges of the fields are
\be
R(X)=2 \ \ \ R(\phi_1)=-1 \ \ \ R(\phi_2)=3 \ \ \ R(\phi_3)=0 \ \ \ 
R(\phi_4)=2
\ee
As before this theory has a runaway behavior in the large field region, 
and the fields are parametrized as
\be  \label{parametrization2}
X =\frac{f}{2 h \mu}e^{2\alpha} , \ \ \ \ \phi_2 =\sqrt{\mu}
e^{-\alpha} ,\ \ \ \ \ \phi_4 =-\frac{f^2}{2 h\sqrt{\mu^5}}e^{3\alpha}
\ \ \ \ \phi_3=0 \ \ \ \phi_4=0
\ee
Near the origin the classical equations of motion break supersymmetry
at tree level at $\phi_i\!=\!0$. The field $X$ is a classical
pseudomodulus whose stability has to be studied perturbatively.  The
pseudomoduli space is stable if $|\langle X\rangle|<\frac{\mu^2}{4 f
  h}$ and $\frac{\l}{\sqrt{\mu}} < \frac{\mu^{3/2}}{2f}$.
\\
We study the effective potential by expanding it in the adimensional
parameter $\frac{f^2}{\mu^3}$, finding
\be
V_{eff}^{(1)}(X)
=-\frac{3 f^2 \l^2(\l^2 X^2+2 \mu^2)}{2(\l^2 X^2+\mu^2)^{3/2}}
-\frac{6 f^2 h^2 X^2 (h^2 X^4 + 2 \mu^2)}{(h^2 X^4+\mu^2)^{3/2}}
\ee
This perturbative analysis holds if the coupling $\l$ is small
at the mass scale of the chiral field $\phi_3$
\be\label{bound3}
\l^2 \ll \l X
\ee
This requirement imposes that the field $X$ cannot be fixed at the origin,
and $R$-symmetry has to be broken in the non supersymmetric vacuum.
The coupling $\l$ has to be small, and we can expand the potential
in the adimensional parameter $\frac{\l}{\sqrt{\mu}}$.
At the lowest order we found that a minimum exists and it 
is 
\be\label{bound4}
X \sim 2^{1/4}\left( \frac{\mu^2}{h^2}-\frac{9 \sqrt{3}\l^2\mu^2}
{15 \sqrt{3} h^2\l^4+4h^4\mu^2} \right)^{1/4}
\ee
Inserting (\ref{bound4}) in (\ref{bound3}) we find the condition under which
the one loop approximation is valid. In this range we found a 
(meta)stable vacuum at non zero vev for the pseudomodulus.

\section{Conclusions}\label{sezioneE}
We have shown that metastable supersymmetry breaking vacua in three
dimensional WZ models are generic.  Relevant couplings potentially
invalidate the perturbative approximation. Nevertheless, as we have
seen, this problem is removed by the addiction of marginal couplings.

This work may be useful for the analysis of spontaneous supersymmetry
breaking in $3D$ gauge theories. This issue has been investigated in
\cite{Witten:1999ds}-\cite{deBoer:1997kr} as a consequence of brane
dynamics.  A preliminary step towards the study of supersymmetry
breaking in the dual field theory living on the branes appeared in
\cite{Giveon:2009bv}, where the three dimensional ISS mechanism has
been discovered for Yang-Mills and Chern-Simons gauge theories.  The
first class has been deeply studied in \cite{Aharony:1997bx}. The
second class has become more important in the last years, because of
its relation with the $AdS_4/CFT_3$ correspondence. It would be
interesting to generalize the three dimensional ISS mechanism of
\cite{Giveon:2009bv} in theories which admit a Seiberg-like dual
description \cite{Amariti:2009rb},
\cite{Giveon:2008zn}-\cite{Armoni:2009vv}.

Another interesting aspect, that needs a further analysis, is the role
of $R$-symmetry.  In four dimensions supersymmetry breaking is
strongly connected with $R$-symmetry and its spontaneous breaking
\cite{Nelson:1993nf}.  Spontaneous $R$-symmetry breaking is a
sufficient condition for supersymmetry breaking.  In three dimensions
similar results hold but the condition seems stronger.  In fact a
perturbative analysis of supersymmetry breaking is possible only if
$R$-symmetry is broken.

\section*{Acknowledgments}
We thank Luciano Girardello and Alberto Mariotti for early
collaboration on this project, and for many interesting discussions
and useful suggestions.  We would also like to thank 
Ken Intriligator and Silvia Penati for useful comments on the draft.  
A.~A.~ and M.~S.~ are supported in part
by INFN, in part by MIUR under contract 2007-5ATT78-002 and in part by
the European Commission RTN programme MRTN-CT-2004-005104.

\appendix

\section{The bounce action for a triangular barrier in three dimensions}
\label{thelifetime}
\begin{figure}
\begin{center}
\includegraphics[width=10cm]{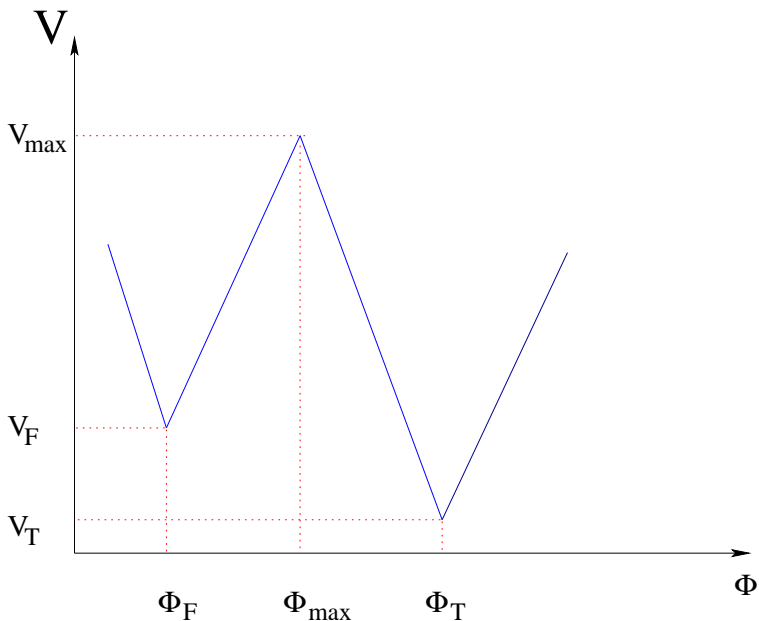}
\caption{Triangular potential barrier} 
\label{barrieratred}
\end{center}
\end{figure}

In this appendix we calculate the bounce action $B$
for a triangular barrier in three dimensions (Figure \ref{barrieratred}).
The bounce action is the difference between the tunneling
configuration and the metastable vacuum in the euclidean
action. The tunneling rate of the metastable state is then
given by $\Gamma=e^{-B}$

Following \cite{duncanjensen} we reduce to the case 
of a single scalar field
with only a false vacuum $\phi_F$ decaying to the true vacuum,
$\phi_T$. The tunneling action is
\be
S_{E}[\phi] = 4 \pi \int_{0}^{\infty}r^2 dr \left(\frac{1}{2}\dot\phi^2
+V(\phi) \right)
\ee
where $\phi(r)$ is the tunneling solution, function of the Euclidean 
radius $r$.
Solving the equation of motion for the $\phi$ field and imposing the
boundary conditions
\bea
&&\lim_{r\rightarrow \infty} \phi(r)=\phi_F \non \\
&&\dot \phi(R_F)=0
\eea
the bounce action is given by
\be\label{bouncaction}
B = S_E[\phi(r)]-S_E[\phi_F]
\ee
where we have subtracted the action for the field sitting at the false vacuum 
$\phi_F$.

It is then helpful to define the gradient of the potential $V'(\phi)$
in terms of the parameters at the extremal points,
\bea \label{gradient}
  \lambda_{F}=&\frac{V_{max}-V_F}{\phi_{max}-\phi_F}& \equiv
  \frac{\Delta V_F}{\Delta \phi_F} \non
  \\ \lambda_{T}=&-\frac{V_{max}-V_T}{\phi_T-\phi_{max}}& \equiv
  -\frac{\Delta V_T}{\Delta \phi_T}
\eea
where the first has positive sign and the second has negative
one.

The solution of the equation of motion at the two side of the
triangular barrier are solved by imposing the boundary conditions, and
a matching condition at some radius $r+R_{max}$, that has to be
determinate.

The choice of the boundary condition proceeds as follows. Firstly
the field $\phi$ reaches the false vacuum $\phi_F$ at a  finite radius
$R_{F}$ and stays there. This imposes
\bea
\phi(R_F)&=&\phi_F\non\\
\dot \phi(R_F)&=&0
\eea
For the second boundary condition we work in the simplified situation
such that the field has initial value $\phi_0<\phi_T$ at radius
$r=0$.  In this way we imposes the conditions
\bea \label{bound2}
\phi(0)&=&\phi_{0}\non\\
\dot \phi(0)&=&0
\eea
%
Solving the equations of motion we have two solution at the two sides
of the barrier
\bea
&&\phi_R(r)=\phi_0-\frac{\lambda_{T}r^2}{6}
~~~~~~~~~~~~~~~~~~~~~~~~~~~~~~~~~~~~~~\text{for~~~~~~} 0<r<R_{max}\\
&&\phi_L(r)=\phi_{max}^{F}-\frac{\lambda_F R_{F}^2}{2}+\frac{\lambda_F
  R_{F}^3}{3r}+\frac{\lambda_F}{6}r^2~~~~~~~~~~\text{for }~~
R_{max}<r<R_F \eea
By matching the derivatives at $R_{max}$ we are able to express $R_F$
as a function of $R_{max}$
\be
R_F^3=(1+c)R_{max}^3
\ee
where $c=-\lambda_T/\lambda_F$. Out of the value of the field at
$R_{max}$ we get two useful relations
\bea
&&\phi_0 = \phi_{max} + \frac{\lambda_T }{6} R_{max}^2\non \\
&&\Delta \phi_{F} = \frac{\lambda_F (3+2 c - 3
  (1+c)^{\frac{2}{3}})}{6} R_{max}^2 \eea
The bounce action $B$ can be evaluated by integrating the equation
(\ref{bouncaction}) from $r=0$ to $r=R_F$. We found
\be \label{bounceuno}
B = \frac{16 \sqrt{6} \pi}{5}
\frac{1+c}{(3+2c-3(1+c)^{2/3})^{3/2}}\sqrt{\frac{\Delta
    \phi_{F}^6}{\Delta V_F}} \ee

\vspace{0.5cm}

There is a second possibility that holds if $\phi_0>\phi_T$.
In this case the field is closed to $\phi_0$ from $R=0$ until a radius 
$R_0$, and then it evolves outside. In this case the boundary conditions 
(\ref{bound2}) become
\bea
\phi(r)~&=&\phi_T  ~~~~~~~~~      0<r<R_T \non \\
\phi(R_T)&=&\phi_T\\
\dot \phi(R_T)&=&0\non
\eea
In this case the equations of motion are solved by
\bea
&&\phi_R(r)=\phi_T~~~~~~~~~~~~~~~~~~~~~~~~~~
~~~~~~~~~~~~~~~~~~~~ \text{for~~~~~~~~}0<r<R_T\non\\
&&\phi_R(r)=\phi_T + \frac{\lambda_T R_{T}^2}{2}-\frac{\lambda_T
  R_T^3}{3r}-\frac{\lambda_T r^2}{6}
~~~~~~~~~~~\text{for~~~~~} R_T<r<R_{max}\\
&&\phi_L(r)=\phi_{F}-\frac{\lambda_F R_{F}^2}{2}+\frac{\lambda_F
  R_{F}^3}{3r}+\frac{\lambda_F r^2}{6}~~~~~~~~~~\text{for }~~
R_{max}<r<R_F\non
\eea
By integrating these solution we can write the bounce action as  
\be \label{bouncedue}
B = \frac{8\pi}{15} \lambda_{F} (R_F^3\Delta\phi_{T} -cR_T^3 \Delta\phi_{F})
\ee
where the relations among the unknowns and the parameters of
the potential are
\bea
&&R_{max}^3 (1+c) = R_F^3+cR_T^3 \non\\
&&\Delta \phi_{T} = \frac{(R_T-R_{max})^2(2 R_T+R_{max})\lambda_T}{6 R_{max}}\\
&&\Delta \phi_{F} = \frac{(R_F-R_{max})^2(2 R_F+R_{max})\lambda_F}{6 R_{max}}
\eea
We conclude by observing that in the limit $R_T=0$ 
(\ref{bouncedue}) coincides with (\ref{bounceuno}).

\section{Coleman-Weinberg formula in various dimensions}
\label{appendixB}
The CW formula for the one-loop superpotential
\be \label{eqn:CW}
  V_{eff}^{(1)} = \frac12 \, \text{STr} \int \frac{d^dp}{(2\pi)^d} \ln
  (p^2+m^2)
\ee
is not always straightforward to compute, since the theory can
contain many fields, and one has to diagonalize the squared mass
matrices.  Some property of the models with metastable vacua can be
analyzed without evaluating the eigenvalues of the squared mass
matrices of component fields of the theory. To this purpose, we
generalize a formula previously given for four-dimensional theories
\cite{Shih:2007av} to work in any dimension.  Indeed, writing
(\ref{eqn:CW}) in spherical coordinates and integrating by parts, we
have
\bea
  V_{eff}^{(1)} &=& \frac{\pi^{d/2}}{\Gamma(d/2)}\, \text{STr}\, \int
  \frac{dp}{(2\pi^d)} \, p^{d-1}\, \ln(p^2+m^2) \non \\ &=&
  -\frac{1}{d} \, \frac{1}{2^{d-1} \, \pi^{d/2} \Gamma(d/2)} \, \text{STr}\,
  \int dp \, \frac{p^{d+1}}{p^2+m^2}
\eea
where $A_d = 2 \pi^{d/2}/ \Gamma(d/2)$ is the d-dimensional spherical
surface. By substituting $d=3$ we recover (\ref{CWa3}).

\end{document}